\documentclass [12pt]{article}
\usepackage[T1]{fontenc}
\usepackage[final]{epsfig}
\usepackage{amssymb}
\usepackage{multicol}
\usepackage{graphics}
\usepackage{color}
\usepackage{ntimes}

\newcommand{\BE}{\begin{equation}}
\newcommand{\EE}{\end{equation}}

\begin{document}

\title{A Markov process associated with plot-size distribution in
Czech Land Registry and its number-theoretic properties}
\author{Pavel Exner$^{1,2}$ and Petr \v Seba$^{2,3,4}$}

\maketitle

\begin{center}

\small{$^{1}$Nuclear Physics Institute, Czech Academy of Sciences,
25068 \v{R}e\v{z} near Prague, Czech Republic}

\small{$^2$ Doppler Institute for Mathematical Physics and Applied
Mathematics, Czech Technical University, B\v{r}ehov\'{a}~7, 11519
Prague~1, Czech Republic}

\small{$^3$ Institute of Physics, Czech Academy of Sciences,
Cukrovarnick\'{a}~10, Prague 8, Czech Republic}

\small{$^4$ Department of Mathematical Physics, University of
Hradec Kr\'alov\'e, V\'{\i}ta Nejedl\'{e}ho~573, Hradec
Kr\'alov\'e, Czech Republic}

\end{center}

\normalsize

\maketitle

\begin{abstract}
The size distribution of land plots is a result of land allocation
processes in the past. In the absence of regulation this is a
Markov process leading an equilibrium described by a probabilistic
equation used commonly in the insurance and financial mathematics.
We support this claim by analyzing the distribution of two plot
types, garden and build-up areas, in the Czech Land Registry
pointing out the coincidence with the distribution of prime number
factors described by Dickman function in the first case.
\end{abstract}

The distribution of commodities is an important research topic in
economy -- see \cite{chat} for an extensive literature overview.
In this letter we focus on a particular case, the allocation of
land representing a non-consumable commodity, and a way in which
the distribution is reached. Generally speaking, it results from a
process of random commodity exchanges between agents in the
situation when the aggregate commodity volume is conserved, in
other words, one deals with pure trading which leads to commodity
redistribution.

Models of this type were recently intensively discussed \cite{sca}
and are usually referred to as kinetic exchange models. Our
approach here will be different, being based on the concept known
as perpetuity. The latter is a random variable $D$ that satisfies
a stochastic fixed-point equation the form
 % ------------- %
\begin{equation}\label{mapping}
    D \triangleq  a(D+1)\,,
\end{equation}
 % ------------- %
where $a$ and $D$ are independent random  variables and the symbol
$\triangleq$ means that the two sides of the equation have the
same probability distribution; by an appropriate scaling, of
course, the value \emph{one} in (\ref{mapping}) can be replaced by
any fixed positive number. It is supposed that the distribution
$P(a)$ of the variable $a$ is given and one looks for the
distribution $Q(D)$ of $D$.

The equation (\ref{mapping}) has a solution provided the variable
$a$ satisfies $\mathrm{mean}(\log a)<0$, cf. \cite{ver}. It
appears in the literature under various names depending on the
field of application; it is known as the Vervaat perpetuity
\cite{ver}, stochastic affine mapping, random difference equation,
stochastic fix point equation, and so on. Before proceeding
further, let us remark that the equation (\ref{mapping}) looks
innocent but it is not. The situation when $a$ is a Bernoulli
variable is tricky, in particular, it was proved in \cite{baron}
that the probability measure associated with $Q(D)$ is singularly
continuous in this case.

Perpetuities themselves appear in different contexts. In the
insurance and financial mathematics, for instance, a perpetuity
represents the value of a commitment to make regular payments
\cite{Goldie}. Another situation where we meet perpetuities arises
in connection with recursive algorithms such as the selection
procedure \emph{Quickselect} -- see, e.g., \cite{Hwang} or
\cite{Mahmoud}, and they also describe random partitioning
problems \cite{hui}.

Related quantities emerge, however, even in purely
number-theoretic problems, in particular, in the probabilistic
number theory they describe the largest factor in the prime
decomposition of a random integer --- see, for instance,
\cite[Cor.~2]{donn} or \cite{Hild, wolf} and references therein.
Specifically, the following claim is valid:
\\ [0.5em]
\textbf{Proposition:} Denote by $p(D,X)$ the probability that a
random integer $n\in(1,X)$ has its greatest prime factor $\leq
X^{1/D}$. The limit $Q(D)=\lim_{X\to\infty}p(D,X)$ exists and
coincides with the solution of the equation $D \triangleq a (D+1)$
corresponding to the uniform distribution, $P(a)=1$.
\\ [0.5em]
Recall that the respective $Q(D)$ is known in the number theory as
the Dickman function \cite{banks}.

Let us pass now to our main subject which is the land plot
distribution. We observe that the present sizes of the plots
result from repeated land redistribution --- land purchases and
sales --- in the past which represents a complex allocation
process.

In attempt to understand it within a simple model, consider first
a situation where the overall area is fixed and there are only
three land owners; one can think about a small island having just
three inhabitants. We consider a discrete time and denote by
$D_j(n),\, j=1,2,3,$ the area of the land owned by the respective
holder at time $t=n$ assuming that the  overall area is equal to
one. Consequently, the triple $\{D_k(n):\: k=1,2,3\}$ belongs for
all $n=1,2,..$ to a 3-simplex, $D_1(n)+ D_2(n)+D_3(n)=1$. The land
trading on the island proceeds as follows: two holders $j,k$ with
$j,k \in \{1,2,3\}$ such that $j\neq k$ are picked randomly and
they trade their lots according to the rule
 % ------------- %
\begin{eqnarray}
\label{purch_sale}
\nonumber
  D_j(n+1) &=& a_n [D_j(n)+D_k(n)] \\
  D_k(n+1) &=& (1-a_n)[D_j(n)+D_k(n)]
\end{eqnarray}
 % ------------- %
where $a_n\in (0,1)$ are independent equally distributed random
numbers. (We suppose that no land exchange involves all the three
holders simultaneously.) Let us take for simplicity $j=1$ and
$k=2$. The simplex condition gives $D_1(n)+D_2(n) =1-D_3(n)$, thus
the relation (\ref{purch_sale}) implies $D_1(n+1)=a_n (1-D_3(n)$.
In a steady situation the areas $D_k$ posses identical
distributions equal to the distribution of the same random
variable $D$; the replacement of $D_1$ and $D_3$ by independent
copies of $D$ then leads to the equation $D\triangleq a(1-D)$. A
simple substitution $D\rightarrow-D$ and $a\rightarrow -a$ leads
finally to the equation (\ref{mapping}), however, the distribution
of $D$ is, of course, invariant under such a transformation, up to
a mirror image. Consequently, the land trading on our island with
three inhabitants leads formally to the lot area distribution
described by the perpetuity equation (\ref{mapping}). Notice that
the constant $1$ appearing in it comes from the simplex constraint
$D_1+ D_2+D_3=1$; it can be regarded as a manifestation of the
fact that the overall area is preserved in the trading.

There is another argument that leads to the same equation and which
can be applied to the case of numerous land traders. The plot size
$D(n)$ owned at the instant $n$ by a chosen one of them can be
regarded as a result of two independent actions. The first step is
the sale of a random fraction $a_n$ of the property owned at time
$n\!-\!1$, i.e. $\:D_{n-1}\rightarrow a_nD_{n-1}$. The following
step is the purchase of a new piece of land of a size $d_n$ and
adding it to the one mentioned. In combination, these actions give
 % ------------- %
\begin{equation}\label{purch_sale2}
    D_n = a_nD_{n-1}+d_n\,, \quad n=1,2,\dots\,.
\end{equation}
 % ------------- %
The process is obviously of Markov type and the distribution of
the plot sizes to which it converges is given by the equation
$D\triangleq aD+d$ where $d$ is the random variable associated
with the acquisitions. In fact, convergence of the process is
closely related to the existence of solution to this equation
\cite{ver}. The process is linked to the Dickman function in case
that the two random variables coincide, $d_n=a_n$ for all
$n\in\mathbb{Z}_+$, and they are identically distributed; then
(\ref{purch_sale2}) obviously leads to (\ref{mapping}). From the
point of land plot reallocations such an assumption is an
idealization and it is formally exact -- as mentioned above -- for
three traders only. It is an open question to what extent this
distribution changes in situations with a larger number of players
involved.

To understand better that how the Dickmann function can arise in
the process, we observe that the solution of (\ref{mapping})  is
obtained formally as the following infinite sum \cite{ver},
 % ------------- %
\begin{equation}\label{sum}
    D=\sum_{n=1}^\infty \prod_{k=1}^n a_1a_2...a_k\,,
\end{equation}
 % ------------- %
where $a_1,a_2,...,a_k,...$ are independent uniformly distributed
random variables. On the other hand, the Markov process
(\ref{purch_sale2}) leads in our particular case, $d_k=a_k$, to
 % ------------- %
 \begin{equation}\label{sum_markov}
    D_{n+1}=a_n+a_na_{n-1}+
    a_na_{n-1}a_{n-2}+..+a_na_{n-1}a_{n-2}...a_1D_1\,,
\end{equation}
 % ------------- %
where $D_1$ is the initial holding of the trader which we put
equal to one. We may naturally relabel the variables and to write
the right-hand side of (\ref{sum_markov}) also as
$a_1+a_1a_2+\cdots$; it is crucial that this leads to the same
random variable $D_{n+1}$ since all the $a_k$'s are independent
and equally distributed. In this form the relation between
(\ref{sum}) and (\ref{sum_markov}) is clearly seen; the question
is whether the two quantities are close to each other in the
situation we are interested in\footnote{Using a modified random
variable we can rewrite (\ref{purch_sale2}) also in alternative
forms, say, $D_n = a_n(D_{n-1}+c_{n-1})$. Some mathematical
results about convergence of such processes with specific random
variables $c$ can be found in \cite{duf}.}. The point is that the
number of the trading steps is of course finite and not very
large. Even without big historical disturbances we can hardly
expect the free land trading to have a history longer than roughly
three centuries. Assuming that a given plot is traded once in a
generation we can thus run the process (\ref{sum_markov})
realistically up to $n\lesssim 10$.

Luckily enough for us the convergence of (\ref{sum_markov}) to
(\ref{sum}) is rather fast: using the mentioned relabelling we
find
 % ------------- %
\begin{equation}\label{convergence}
    D-D_{n+1}=a_1a_{2}a_{3}...a_n \widetilde{D}\,,
\end{equation}
 % ------------- %
where $D$ and $\widetilde{D}$ are statistically equivalent.
Denoting as usual by $\mathbb{E}(a)$ the mean of $a$, we get
therefore
 % ------------- %
\begin{equation}\label{mean}
    \mathbb{E}(D-D_{n+1})=\mathbb{E}(a)^n
    \mathbb{E}(\widetilde{D})=2^{-n}\,,
\end{equation}
 % ------------- %
since $a$ is identically distributed in $(0,1)$ by assumption and
$\mathbb{E}(\widetilde{D})=1$, which means that the convergence is
exponentially fast. But this is not all, one can also show that
the convergence is extremely shape robust. Indeed, take the Markov
rule $D_{n+1}=a_n(D_n+1)$ and denote by $G_n(t)$ the probability
that $D_n<t$.  For the uniformly distributed variable $a$ we then
have
 % ------------- %
\begin{equation}\label{prob1}
G_{n+1}(t)=\int_0^1 G_n\left(\frac{t}{a}-1\right) \mathrm{d}a\,,
\end{equation}
 % ------------- %
and consequently, the densities $g_n(t):=G_n'(t)$ satisfy
 % ------------- %
\begin{equation}\label{prob2}
g_{n+1}(t)=\int_0^1 g_n\left(\frac{t}{a}-1\right)
\frac{\mathrm{d}a}{a}\,;
\end{equation}
 % ------------- %
the support of all these functions lies by definition of the
nonnegative real axis, $G_n(t)=g_n(t)=0$ for $t<0$. A simple
substitution $u=t/a-1$ then gives finally a relation between the
densities $g_{n+1}$ and $g_n$, namely
 % ------------- %
\begin{equation}\label{prob3}
g_{n+1}(t)=\int_{t-1}^\infty g_n(u) \frac{\mathrm{d}u}{u+1}\,.
\end{equation}
 % ------------- %
Let us now start with the situation far from the expected
equilibrium assuming, for instance, that all the owners have at
the initial instant land plots of the same are, i.e.
$g_1(t)=\delta(t-1)$; then (\ref{prob3}) gives
 % ------------- %
 $$
    g_2(t)= \left\{
\begin{array}{lcl}
  \frac12 & \quad \mathrm{if}\;\; t\in [0,2) \\ [2pt]
  0 & \quad \mathrm{if}\;\; t>2
\end{array} \right.
 \qquad
    g_3(t) = \left\{
\begin{array}{lcl}
  \frac{1}{2}\ln 3 & \quad \mathrm{if}\;\; t\in [0,1] \\ [2pt]
  \frac{1}{2}\ln\left(\frac{3}{t}\right) & \quad \mathrm{if}\;\; t\in [1,3]
  \\[2pt]
  0 & \quad \mathrm{if}\;\; t>3
\end{array} \right.
 $$
 % ------------- %

\noindent and so on. It can be seen easily from (\ref{prob3}) that
the functions $g_n(t), n=3,4,\cdots$, have the following
properties: $g_n(t)=c_n$ for $t\in (0,1)$ where $c_n$ is a
constant depending on $n$ only, and moreover, $g_n(t)$ is
decreasing for $t\in [1,n]$ and $g_n(t)=0$ holds for $t>n$.
Furthermore, we have $c_n\to \mathrm{e}^{-\gamma}$ as $n\to
\infty$ with $\gamma$ being the Euler--Mascheroni constant ---
note that $\mathrm{e}^{-\gamma}$ is the value of the Dickman
function for $t\in (0,1)$. Hence even if the original distribution
had nothing in common with the Dickman function, the densities
$g_n(t)$ are form robust and approach rapidly such a shape. In
fact integrating further we find that the densities $g_n$ for
$n>3$ are already very close to the Dickman limit $g_\infty$.

The conclusion for our model is that there is a chance to see the
equilibrium situation in the land plot distributions provided the
trading go undisturbed for at least four generations. If this is
the case it makes sense to ask about a relation between the plot
distribution and the equation (\ref{mapping}); it is clear that
only an inspection of actual data can show whether such a model
assumption is good or not.

Let us thus look whether these considerations have something in
common with land plot distribution in reality. One has to be
cautious, of course, when choosing which types of plots are to be
considered. Recall, for instance, that A.~Abul-Magd tried recently
to describe the wealth distribution in the ancient Egyptian
society using areas of the house found by the excavations in Tell
el-Amarna \cite{abulmagd}. Their distribution exhibited a Pareto
like distribution \cite{pareto} known to describe the wealth
allocation among individuals. It has an algebraic tail, and
therefore it behaves in a way different from the Vervaat
perpetuity for constant $P(a)$; recall that the Dickman function
$Q(D)$, for instance, satisfies asymptotically the inequality
$Q(D)<D^{-D}$ for large $D$. It is not surprising, however, that
this case does not fit into our scheme, the aggregate volume being
not locally conserved: upgrading a house as a result of one's
wealth need not affect the areas of the neighbouring houses.

The land plots we look for have to satisfy several criteria. As
the above example suggests, they have to be arranged in connected
areas, so that the gain of a purchaser is the same as the loss of
the corresponding seller. At the same time, they must be divisible
so one can sell and buy parts of them. Choosing such a plot type,
one can look into the land registry where the present holdings are
recorded. As we have said they are the result of repeated land
purchase and land sell done by the ancestors in the past, but we
are not going to look into the history being interested in the
resulting distribution. We have to make sure, however, that the
process was not affected by the outside influences like agrarian
reforms or other forms of redistribution \emph{en gros}.

\begin{figure}
\begin{center}
  \includegraphics[height=9cm,width=15cm]{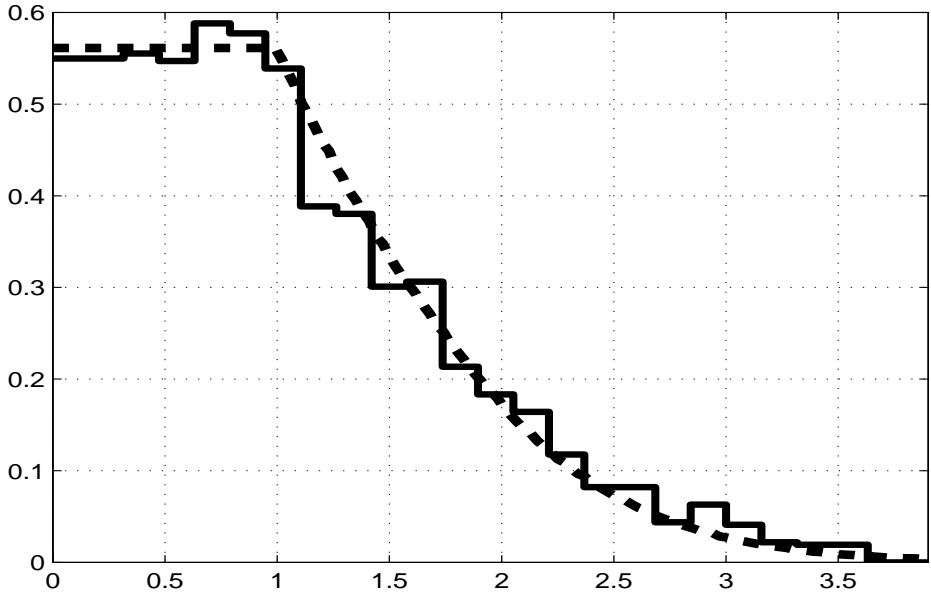}\\
\end{center}
  \caption{
The probability density of the normalized garden areas (full line)
is compared with the Dickman function (dashed line). We have used
areas of 4000 gardens located in the two towns.
  }
  \label{dic}
\end{figure}

One plot type suitable for our purpose are gardens in urban areas.
We used the Czech real estate cadastre concentrating on the sizes
of four thousand gardens in the urban area of the towns
\emph{Rychnov nad Kn\v{e}\v{z}nou} and \emph{Dobru\v{s}ka} in East
Bohemia. To compare their distribution with the perpetuity result
mentioned above we need, of course, a proper normalization: we
choose the scale in which the mean size of the plot is equal to
one. The result confirms our conjecture: the probability
distribution of the garden areas coincides with the Dickman
function as shown in Figure~\ref{dic}.

This finding can be easily understood. Gardens in urban areas are
desirable properties, and as a result, any piece of a garden is
equally good for the market. This means the process
(\ref{purch_sale}) goes on with the variables $a$ and $d$
approximately homogeneously distributed over the interval $(0,1)$,
so that the Dickman function gives a good fit confirming our
expectation. Another conclusion we can make without looking into a
detailed history is that the overall area of gardens in these
towns did not change significantly in the course of the time.

Another suitable land plot type recorded in the cadaster which was
not affected by agrarian reforms are yards and the build-up areas.
In the latter case we take into account the areas only without
paying attention to the type (number of floors, etc.) of the
buildings which may be constructed on them. The situation is
somewhat different from the garden distribution case, since now we
need not restrict ourselves to the urban areas.  This enables us
to work with a much larger data set; altogether we employed data
about 47000 yards and build-up areas.

\begin{figure}
\begin{center}
  \includegraphics[height=9cm,width=15cm]{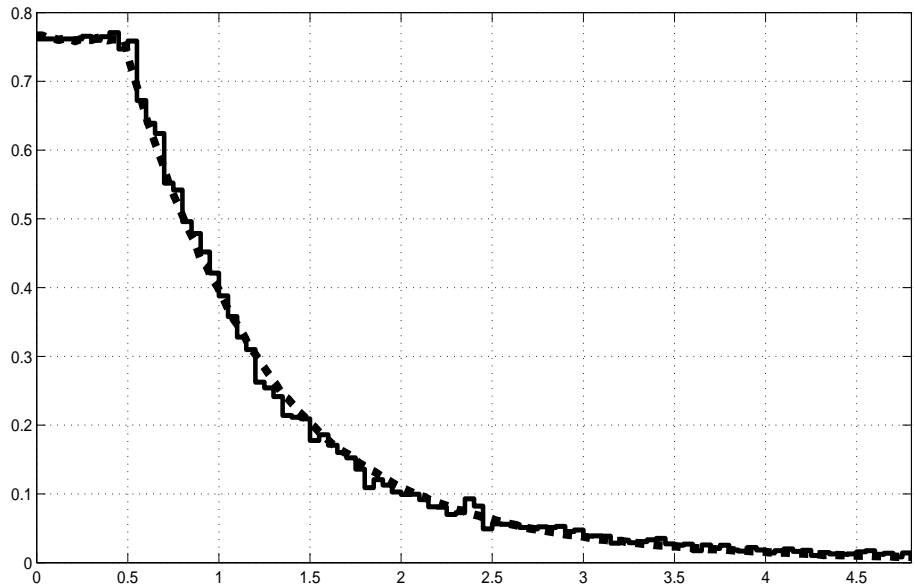}\\
\end{center}
  \caption{
The probability density of the normalized sizes of the yard and
build-up areas (full line) compared with the solution of eq.
(\ref{mapping}) with  uniformly distributed $a\in (0,1.52)$
(dashed line). We have used the sizes of 47000 yards and build-up
areas.
  }
  \label{dic2}
\end{figure}

The allocation process is different because it is not conservative
in this case. In the course of time a new size of build-up area or
yard can be a result of merging a number of smaller areas into a
larger one, and at the same time, a larger area can come from the
transformation of another land type into the building land. To
describe such a process we use again the equation (\ref{mapping})
with the uniformly distributed variable $a$ supposing that all
changes occur with equal probability, however, we change the
variable range taking $a\in (0,A)$ with $A>1$ to take into account
the fact that the build-up area can expand. The result is plotted
on the figure \ref{dic2} and we see that choosing $A=1.52$ we get
an excellent fit.

It would be interesting to test the concept described here on
other plot types such as the distribution of fields or forests.
For this purpose, unfortunately, the Czech land registry is not
suitable because in this case the distribution was formed not only
by standard market forces but also by processes like
collectivization, etc. A brief look at the corresponding data
shows that the distributions cannot be described with the help of
(\ref{mapping}) with a symmetric distribution $P(a)$.

{\bf Acknowledgement:} We thank the referees for useful remarks.
The research was supported by the Czech Ministry of Education,
Youth and Sports within the project LC06002. The cooperation of
the Land Registry District Office in Rychnov nad Kn\v{e}\v{z}nou
is gratefully acknowledged.


\begin{thebibliography}{99}

\bibitem[AM02]{abulmagd}
A.Y.~Abul-Magd: Wealth distribution in an ancient Egyptian
society, \emph{Phys. Rev.} \textbf{E66} (2002), 057104 .

\bibitem[BR01]{baron}
M.~Baron, A.L.~Rukhin: Perpetuities and asymptotic change-point
analysis, \emph{Stat. and Probab. Lett.} \textbf{55} (2001),
29--38.

\bibitem[BS07]{banks}
W.D.~Banks, I.E.~Shparlinski: Integers with a large smooth
divisors, \emph{Integers} \textbf{7} (2007), A17; see also
\texttt{arXiv:math.NT/0601460}

\bibitem[CC07]{chat}
A.~Chatterjee, B.K.~Chakrabarti: Kinetic exchange models for
income and wealth distributions, \texttt{arXiv:0709.1543
[physics.soc-ph]}

\bibitem[DG93]{donn}
P.~Donnelly, G.~Grimmett: On the asymptotic distribution of large
prime factors, \emph{J. Lond. Math. Soc.} \textbf{47} (1993),
395--404.

\bibitem[Du90]{duf}
D.~Dufresne: The distribution of a perpetuity with application to
risk theory and pension funding, \emph{Scand. Actuarial J.}
(1990), 750--783.

\bibitem[GM00]{Goldie}
Ch.M.~Goldie, R.A.~Maller: Stability of perpetuities, \emph{Ann.
Probab.} \textbf{28} (2000), 1195--1218.

\bibitem[HT93]{Hild}
A.~Hildebrand, G.~Tenebaum: Integers without large prime factors,
{\em J. de Th\'eorie des Nombres de Bordeaux} {\bf 5} (1993),
411--484.

\bibitem[Hu05]{hui}
T.~Huillet: Random partitioning problems involving Poisson point
processes o the interval, \emph{Int. J. Pure Appl. Math.}
\textbf{24} (2005), 143--179.

\bibitem[HT02]{Hwang}
H.--K.~Hwang, T.--H.~Tsai: Quickselect and the Dickman function,
\emph{Combin. Probab. Comput.} \textbf{11} (2002), 353--371.

\bibitem[MMS95]{Mahmoud}
H.~Mahmoud, R.~Modarres, R.~Smythe:  Analysis of QUICKSELECT: an
algorithm for order statistics, \emph{RAIRO Inform. Theor. Appl.}
\textbf{29} (1995), 255-–276.

\bibitem[Pa1897]{pareto}
V. Pareto: \emph{Course d'Economie Politique}, Macmillan, Paris
1897.

\bibitem[SGG06]{sca}
E.~Scalas, M.~Gallegati, E.~Guerci, D.~Mas, A.~Tedeschi: Growth
and allocation of resources in economics: The agent-based
approach, \emph{Physica} \textbf{A370} (2006), 86--90.

\bibitem[Ve79]{ver}
W.~Vervaat: On a stochastic difference equation and a
representation of of non-negative infinitely divisible random
variables, \emph{Adv. Appl. Prob.} \textbf{11} (1979), 750--783.

\bibitem[MV07]{wolf}
E.W.~Weisstein: Dickman Function, from \emph{MathWorld -- A
Wolfram Web Resource,
http://mathworld.wolfram.com/DickmanFunction.html}.


\end{thebibliography}
\end{document}